\documentclass[preprint]{aastex}
\slugcomment{draft of \today}
\shorttitle{FRBs as Cosmic Magnetic Field Probes}
\shortauthors{Akahori et al.}

\usepackage{color}

\begin{document}
\title{Fast Radio Bursts as Probes of Magnetic Fields in the Intergalactic Medium}

\author{
Takuya Akahori\altaffilmark{1,2}, Dongsu Ryu\altaffilmark{3,4,7} and B. M. Gaensler\altaffilmark{5,6}
}

\altaffiltext{1}{Graduate School of Science and Engineering, Kagoshima University, Kagoshima 890-0065, Japan; akahori@sci.kagoshima-u.ac.jp}
\altaffiltext{2}{SKA Organisation, Cheshire SK11 9DL, UK}
\altaffiltext{3}{Department of Physics, UNIST, Ulsan 44919, Korea; ryu@sirius.unist.ac.kr}
\altaffiltext{4}{Korea Astronomy and Space Science Institute, Daejeon, 34055, Korea}
\altaffiltext{5}{Dunlap Institute for Astronomy and Astrophysics, The University of Toronto, Toronto ON M5S 3H4, Canada; bgaensler@dunlap.utoronto.ca}
\altaffiltext{6}{Sydney Institute for Astronomy, School of Physics, The University of Sydney, NSW 2006, Australia}
\altaffiltext{7}{Author to whom all correspondence should be addressed.}

\keywords{intergalactic medium --- large-scale structure of universe --- magnetic fields --- polarization --- radio continuum: general}

\begin{abstract}
We examine the proposal that the dispersion measures (DMs) and Faraday rotation measures (RMs) of extragalactic linearly-polarized fast radio bursts (FRBs) can be used to probe the intergalactic magnetic field (IGMF) in filaments of galaxies. The DM through the cosmic web is dominated by  contributions from the warm-hot intergalactic medium (WHIM) in filaments and from the gas in voids. On the other hand, RM is induced mostly by the hot medium in galaxy clusters, and only a fraction of it is produced in the WHIM. We show that if one excludes FRBs whose sightlines pass through galaxy clusters, the line-of-sight strength of the IGMF in filaments, $B_{||}$, is approximately $C(\langle 1+z \rangle/f_{DM})(RM/DM)$, where $C$ is a known constant. Here, {the redshift of the FRB is not required to be known;} $f_{DM}$ is the fraction of total DM due to the WHIM, while $\langle 1+z \rangle$ is the redshift of interevening gas weighted by the WHIM gas density, both of which can be evaluated for a given cosmology model solely from the DM of an FRB. Using data on structure formation simulations and a model IGMF, we show that $C(\langle 1+z \rangle/f_{DM})(RM/DM)$ closely reproduces the density-weighted line-of-sight strength of the IGMF in filaments of the large-scale structure.

\end{abstract}

\section{Introduction}
\label{section1}

The generation and evolution of the intergalactic magnetic field (IGMF) bears on many aspects of astrophysics, yet its real nature is not well understood \citep[see][for review]{rsttw12,wrsst12}. It is anticipated that the Square Kilometre Array and its precursors and pathfinders can explore the IGMF in filaments of galaxies with Faraday rotation measure (RM) \citep{agr14,aktr14,ide14,gae15,joh15,tay15}. RM through filaments has been predicted with cosmological simulations, but the predictions have not yet converged; expected magnitudes are a few to several rad m$^{-2}$ \citep{ar10,ar11} based on the IGMF model of \cite{ryu08}, or smaller in other models \citep[e.g.,][]{vaz14,mar15}.

Since RM is an integral of magnetic field along the line-of-sight (LOS), $B_\parallel$, weighted with electron density, $n_e$, we need to know $n_e$ for the intergalactic medium (IGM) to estimate the strength of the IGMF. Dispersion measure (DM), the free electron column density along the LOS, has been suggested as a possible probe of the IGM density \citep{iok03,ino04}, but can only be measured for the IGM by observation of a bright, brief, radio transient at cosmological distances. Fast radio bursts (FRBs) are a new phenomenon which appear to indeed provide us with these measurements of extragalactic DMs \citep{lor07,kea12,lor13,tho13,tot13,zha13,kas13,pet15,mac15,ma15a,cha15,kea16,spitler16}.

A number of FRBs have now been reported \citep{pbj16},\footnote{FRB
Catalogue, http://astronomy.swin.edu.au/pulsar/frbcat/, version 1.0} with
DMs in the range $\sim 400$ -- $1600$ ${\rm pc~cm^{-3}}$. These large DMs
imply that FRBs occur at cosmological redshifts, $z\sim 0.5$ -- $1$
\citep{tho13,dolag15}. \citet{ma15b} have reported the first detection of
linear polarization in a FRB: for FRB~110523, \citet{ma15b} find DM $=
623.3$ ${\rm pc~cm^{-3}}$ and RM $=-186$ ${\rm rad~m^{-2}}$, and conclude
that the RM was induced in the vicinity of the source itself or within the
host galaxy.  {\citet{kea16} have claimed\footnote{Note that
this claim is under scrutiny; see \cite{wb16}.} an identification of host
elliptical galaxy at $z = 0.492 \pm 0.008$ for FRB 150418 with DM $= 776.2
\pm 0.5$ ${\rm pc~cm^{-3}}$ and RM $= +36 \pm 52$ ${\rm rad~m^{-2}}$}.
{\citet{spitler16} and \citet{scholz16} have presented
observations of repeating bursts for FRB 121102, suggesting that the source
object could be a young neutron star.}

The DMs and RMs of extragalactic linearly-polarized FRBs together may be used to explore the IGMF. For an extragalactic source located at $z=z_i$, these quantities can be written at the observer's frame as 
\begin{equation}\label{eq:DM}
DM = C_{\rm D}\int_{0}^{z_i}\frac{n_{\rm e}(z)}{(1+z)}\frac{dl(z)}{dz}dz~~{\rm pc~cm^{-3}},
\end{equation}
\citep[e.g.,][]{mc14,dz14} and 
\begin{equation}\label{eq:RM}
RM = C_{\rm R}\int_{z_i}^{0}\frac{n_{\rm e}(z) B_\parallel(z)}{(1+z)^{2}}\frac{dl(z)}{dz}dz~~{\rm rad~m^{-2}}, \\
\end{equation}
\citep[e.g.,][]{ar11}, respectively. Here, $n_{\rm e}(z)$ is the proper electron density in the cosmic web at a redshift $z$ in units of ${\rm cm^{-3}}$, $B_\parallel(z)$ the LOS component of the IGMF at $z$ in $\mu$G, and $dl(z)$ the LOS line element at $z$ in kpc, with the numerical constants having values $C_{\rm D} \simeq 1000$ and $C_{\rm R}\simeq 811.9$. Traditionally, the LOS magnetic field strength is estimated as
\begin{eqnarray}\label{eq:RMDM}
B_{||}^\dagger &=& \frac{C_{\rm D}RM}{C_{\rm R}DM} \nonumber\\
 &=& 12.3\left(\frac{RM}{10~{\rm rad~m^{-2}}}\right)\left(\frac{DM}{10^3~{\rm pc~cm^{-3}}}\right)^{-1}~{\rm nG}.
\end{eqnarray}
In this paper, we will show that the above method needs to be revised in the cosmological context.

The idea of using the DMs and RMs of FRBs to probe the IGMF was previously presented by \citet{zhe14}. They employed simple analytic models of the IGM and IGMF and did not consider cosmic web structures. In this paper, using the results of cosmological structure formation simulations and a model IGMF based on a turbulent dynamo in the large-scale structure (LSS) of the universe, we quantify the contribution of the cosmic web to the DMs and RMs of FRBs. We investigate how $B_{||}^\dagger$ in equation (\ref{eq:RMDM}) compares with the IGMF strength in filaments, and propose a modified formula. We do not consider other contributions to DM and RM, such as those from host galaxies or local environments of FRBs or from the foreground Milky Way  \citep[see, e.g.,][and also \S4 below for discussions of those contributions]{agr14,dolag15,ma15b,kon15}. The rest of this paper is organized as follows: the models and calculation are described in \S2, the results are shown in \S3, and the discussion and summary are set out in \S4.

\section{Models and Calculation}
\label{section2}

The models adopted in this paper are essentially the same as those of
\citet{ar10,ar11}. The LSS of the universe is represented by the data of
$\Lambda$CDM universe simulations with $\Omega_{\rm b0}=0.043$, $\Omega_{\rm
m0}=0.27$, $\Omega_{\rm \Lambda 0}=0.73$, $h\equiv H_0/(100~{\rm
km/s/Mpc})=0.7$, $n=1$, and $\sigma_8=0.8$. The simulation box has a $(100\
h^{-1}{\rm Mpc})^3$ volume including $512^3$ uniform grid zones for gas and
gravity and $256^3$ particles for dark matter. Sixteen simulations with
different realizations of initial conditions were used to compensate for
cosmic variance. For the IGMF, we assume that turbulence is generated during
the formation of LSS, and that the magnetic field is produced as a
consequence of the amplification of weak seeds by turbulent flow motions.
The strength of our model IGMF for the warm-hot intergalactic medium (WHIM) in filaments is of order $\langle B\rangle \sim 10$ nG or $\langle \rho B\rangle/\langle \rho\rangle \sim 100$ nG at $z=0$ \citep[see][for details]{ryu08}.

{The cosmic space from redshift $z = 0$ to $z=5$ for our
calculation has been reconstructed using simulation outputs at $z_{\rm out}
= $ 0, 0.2, 0.5, 1.0, 1.5, 2.0, 3.0, and 5.0, following the usual method of
cosmological data stacking \citep[e.g.,][]{sblt00}. A total of 56 simulation
boxes were stacked to reach $z = 5$, and the boxes nearest a given redshift
were used for that redshift. The stacked boxes were randomly selected from
sixteen simulations and then randomly rotated to avoid any artificial coherent structure along the LOS.} Observers were placed at the center of galaxy groups to reproduce the environment of the Milky Way; we chose the galaxy groups that have an X-ray emissivity-weighted temperature similar to that of the Local Group, 0.05 keV $\leq$ $k T_X$ $\leq$ 0.15 keV \citep[see][for details]{ar11}.

Our calculation covers a $20^\circ \times 20^\circ$ field-of-view (FOV) with $400 \times 400$ pixels. The corresponding spatial resolution is $0\hbox{$.\!\!^\circ05$}$, which would be sufficient to resolve major structures of density and magnetic field in the cosmic web. We produced 100 realizations of the FOV and put one FRB at the center of each pixel, so the total number of FRB smaples is 16 million. The redshift of FRBs was randomly chosen from the redshift range $ 0 \le z \le 5$. We note that this large number of FRBs is used in our calculation to compensate the cosmic variance and to avoid statistical fluctuation. It does not mean that future observations will need such numbers of FRBs in order to estimate the IGMF strength (see \S 4). We further note that the redshifts of FRBs do not need to be measured in order to conduct the analysis that we now consider.

LOS integrations for the $i$-th FRB were performed from the observer ($z=0$) up to the FRB's redshift ($z=z_i$). In this paper, we present integrals of several quantities. DM and RM are calculated with equations (\ref{eq:DM}) and (\ref{eq:RM}), respectively, and the path length is calculated as,
\begin{equation}
L = \int_{0}^{z_i} \frac{dl(z)}{dz}dz.
\end{equation}
The density-weighted strength of the IGMF, $B$, is
\begin{equation}
B = \int_{0}^{z_i} n_{\rm e}(z)B(z) \frac{dl(z)}{dz}dz {\bigg/} \int_{0}^{z_i} n_{\rm e}(z) \frac{dl(z)}{dz}dz,
\end{equation}
the density-weighted LOS strength of the IGMF, $B_{||}$, is 
\begin{equation}
B_{||} = \int_{0}^{z_i} n_{\rm e}(z)B_{||}(z) \frac{dl(z)}{dz}dz {\bigg/} \int_{0}^{z_i} n_{\rm e}(z)\frac{dl(z)}{dz}dz.
\end{equation}
$B_{||}^\dagger$ is then calculated using equation (\ref{eq:RMDM}),

\begin{deluxetable}{lll}
\tablenum{1}
\tabletypesize{\small}
\tablewidth{0pt}
\tablecaption{
Summary of Notations for the IGM Components
\label{tab:txy}
}
\tablehead{Notation & Target & Criterion}
\startdata
ALL & all gas & \\
T79 & gas in clusters & $T \ge 10^7$ K\\
T57 & gas in filaments & $10^5$~K $\le T < 10^7$ K\\
T45 & gas in sheets & $10^4$ K $\le T < 10^5$ K\\
T04 & gas voids & $T < 10^4$ K\\
TS0 & LOSs avoiding clusters & $T_X < T_X^*$, $S_X < S_X^*$
\enddata
\end{deluxetable}

The integrations were made over the whole cosmic web (labeled ALL), as well
as its components classified with the IGM temperature, $T$.
{Here we adopt the notation T$xy$ to indicate that only gas
with temperature in the range $10^x$~K $\le T < 10^y$~K has been integrated
through LOSs (see Table~\ref{tab:txy}): T79 for hot gas in clusters of
galaxies with $T \ge 10^7$ K, T57 for the WHIM in filaments of galaxies with
$10^5$~K $\le T < 10^7$~K, T45 for gas in possible sheet-like structures
with $10^4$ K $\le T < 10^5$ K, and T04 for gas in voids with $T < 10^4$ K.}

The integrations over different components of the cosmic web cannot be
directly compared with real observations. {To estimate the
IGMF in filaments, we attempted to select LOSs that avoid galaxy clusters
with a criterion based on X-ray surface temperature ($T_X$) and brightness
($S_X$); that is, LOSs for which pixels with $T_X > T_X^*$ and $S_X > S_X^*$
have been excluded. We adopted the TS0 scheme of \citet{ar11} with $T_X^* = 10^7$ K, and $S_X^* = 10^{-10}$ ${\rm erg~s^{-1}~cm^{-2}~sr^{-1}}$, respectively, which mimic a detection limit of X-ray facilities.} The TS0 scheme should eliminate most of the LOSs that go through galaxy clusters \citep[see][]{ar11,agr14}.

In an attempt to accurately extract the IGMF strength from the DM and RM (see \S3.3), we also present the fraction of DM due to different components of the cosmic web, $f_{DM}$, and the density-weighted redshift along LOSs for different components of the cosmic web,
\begin{equation}\label{eq:nz}
\langle 1+z \rangle =
\int_{0}^{z_i}\frac{n_{\rm e}(z)}{(1+z)}\frac{dl(z)}{dz}dz {\bigg/}
\int_{0}^{z_i}\frac{n_{\rm e}(z)}{(1+z)^2}\frac{dl(z)}{dz}dz.
\end{equation}
The specific form of $\langle 1+z \rangle$ is motivated by the density and redshift dependences in equations (\ref{eq:DM}) and (\ref{eq:RM}).

\begin{figure}
\begin{center}
\includegraphics[width=100mm]{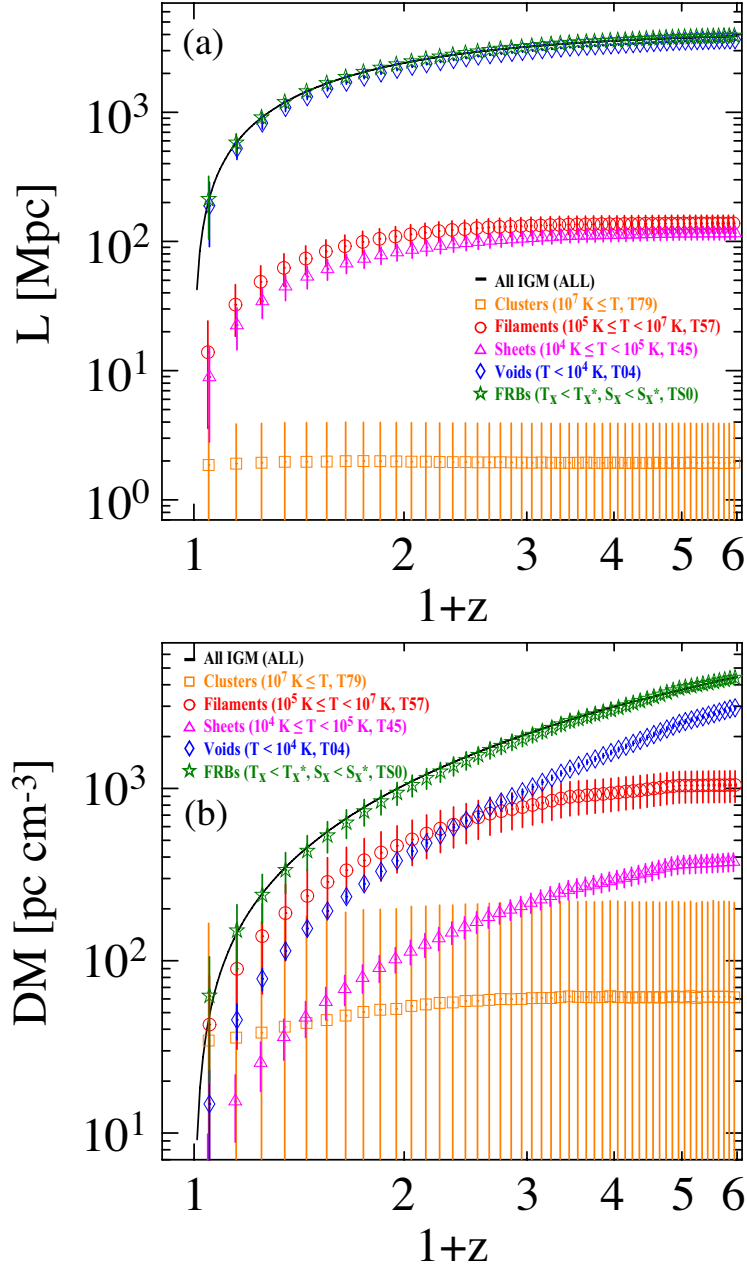}
\end{center}
\vskip -0.6cm
\caption{
The (a) path length ($L$) and (b) dispersion measure (DM) to FRBs,
integrated along LOSs up to the indicated redshift, $z$, for various
components of the cosmic web: {T79 (orange squares) for the
hot gas in clusters of galaxies with $T \ge 10^7$ K, T57 (red circles) for
the WHIM in filaments of galaxies with $10^5$~K $\le T < 10^7$~K, T45
(magenta triangles) for the gas in possible sheet-like structures with
$10^4$ K $\le T < 10^5$ K, T04 (blue diamonds) for the gas in voids with $T
< 10^4$ K, and TS0 (green stars) for LOSs excluding pixels with clusters.} Symbols and error bars represent the average and standard deviation, respectively. The black lines are the analytic solutions for the whole cosmic web (see text).
}\label{f1}
\end{figure}

\placefigure{f1}

We then calculated statistical quantities, the average,
\begin{equation}\label{eq:ave}
X_{\rm avg}(z) = \frac{1}{n_z} \sum_{z-\Delta z/2}^{z+\Delta z/2} X(z_i),
\end{equation}
the standard deviation,
\begin{equation}\label{eq:std}
X_{\rm sd}(z) = \sqrt{\frac{1}{n_z} \sum_{z-\Delta z/2}^{z+\Delta z/2}[X(z_i)-X_{\rm avg}(z)]^2},
\end{equation}
and the root-mean-square (rms),
\begin{equation}\label{eq:rms}
X_{\rm rms}(z) = \sqrt{\frac{1}{n_z} \sum_{z-\Delta z/2}^{z+\Delta z/2} X(z_i)^2},
\end{equation}
where $X$ is one of the integrals we consider. {Here, the
summations are over FRB samples in redshift bins of width $\Delta z=0.1$.}

\section{Results}
\label{section3}

\subsection{Dispersion Measure}
\label{subsection3.1}

We first present results on the DM and path length ($L$) as a function of
redshift. Figure \ref{f1}(a) shows the average and variance of $L$.
Different symbols represent values of $L$ through different components of
the cosmic web and for TS0. The black line indicates $L =
\int_{0}^{z_i}(dl/dz)dz$, which should be identical to the average path
length, calculated numerically for ALL (not shown). The figure indicates
that the path length through the cosmic web is contributed primarily by T04
(voids, blue diamonds) and secondarily by T57 (filaments, red circles) and T45
(sheet-like structures, magenta triangles). While $L$ for T04 continues to
increase with redshift, those for T57 and T45 increase and then converge to
$\sim 100 - 150$ Mpc around $z\sim 2$, since these structures are not yet
fully developed at high redshift. The value of $L$ for T79 (clusters of
galaxies, orange squares) is small and only up to $\sim 2$ Mpc on average;
hence the value of $L$ for TS0 (cluster-subtracted, green stars) is almost the same as that for ALL.

Figure \ref{f1}(b) shows the average and variance of DM. Again, different symbols represent DMs through different components of the cosmic web and for TS0. The black line is the DM calculated analytically for the whole cosmic web using equation (\ref{eq:DM}) with the average cosmic density; it is identical to ${\rm DM}$ calculated numerically for ALL (not shown). The figure demonstrates that the IGM DM of an FRB is dominated by the contributions of T57 and T04. At the lowest redshift ($z \simeq 0.0 - 0.1$), the values of ${\rm DM}$ for T79 and T57 are comparable, although T79 has a large variance depending on the local environment of the observer. At $z \la 1.5$, the value of ${\rm DM}$ is largest for T57, while at higher redshift, the ${\rm DM}$ for T04 dominates. Since the value of ${\rm DM}$ for T79 is small for most of the redshift range, both of the average and standard deviation of ${\rm DM}$ for TS0 is close to those for ALL.

We see that the standard deviation of DM for TS0 (ALL) is small enough, suggesting that the DM can be used to independently estimate the redshift of an FRB once the cosmological model is given. Specifically, the 1$\sigma$ error in DM corresponds to $\sim 2$ redshift bins, i.e., $\delta z \sim 0.2$, for the range of observed DMs for FRBs, $\simeq 400 - 1600$~${\rm pc~cm^{-3}}$. Such a variance is in agreement with previous works \citep[e.g.,][]{dolag15}.
Observed DMs, however, contain contributions from host galaxies of FRBs and the Milky Way, in addition to those from the LSS. This will result in a systematic overestimation/error in the redshift estimation. We will revisit this issue in \S 4.

\begin{figure}
\begin{center}
\includegraphics[width=100mm]{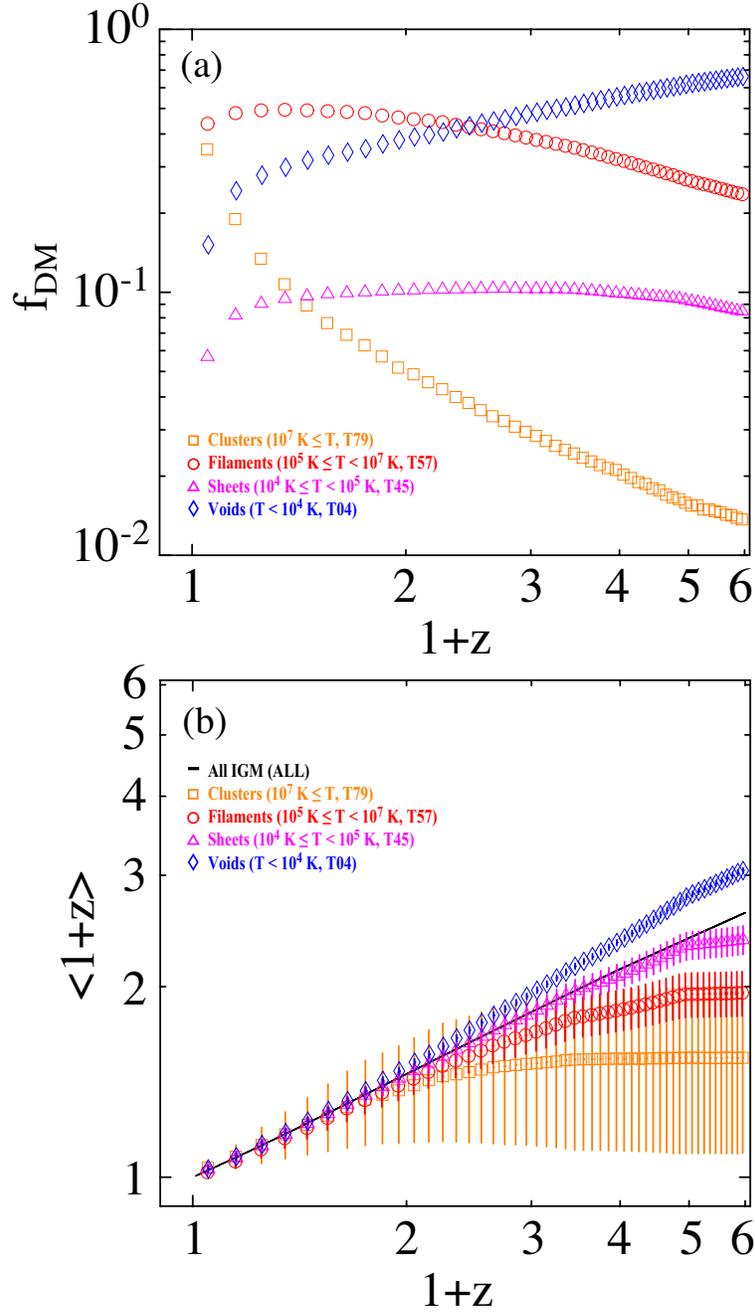}
\end{center}
\vskip -0.6cm
\caption{
(a) The average DM fraction, $f_{DM}$, for T79 (orange squares), T57 (red
circles), T45 (magenta triangles), and T04 (blue diamonds), calculated by normalizing ${\rm DM}$ for each component by the value of ${\rm DM}$ for ALL, integrated along LOSs up to a redshift $z$. (b) As for panel (a), but showing the variation of the density-weighted redshift $\langle 1+z \rangle$ with redshift. Symbols and error bars in (b) represent the average and standard deviation, respectively. The black line in (b) is the analytic solution for the whole cosmic web (see text).
}\label{f2}
\end{figure}

\placefigure{f2}

Figure \ref{f2}(a) shows the average fraction of DM, $f_{DM}$, contributed by different components of the cosmic web, i.e., ${\rm DM}$ for a given component normalized by ${\rm DM}$ for ALL. The value of $f_{DM}$ for T57 is $\sim 40$ -- 50~\% at $z \la 1.5$ and decreases to $\sim 20~\%$ at $z\sim 5$, while $f_{DM}$ for T04 increases from $\sim 20$ \% to $\sim 70$ \% as we move from low to high redshifts. The values of $f_{DM}$ for T79 and T45 are small, with $f_{DM} \la 10$ \% except for T79 at $z \la 0.4$.

Figure \ref{f2}(b) shows the average and variance of the density-weighted redshift, $\langle 1+z \rangle$, in equation~(\ref{eq:nz}). Different symbols represent values of $\langle 1+z \rangle$ through different components of the cosmic web. The black line is the value of $\langle 1+z \rangle$ calculated analytically for the whole cosmic web, which approximates to $\langle 1+z \rangle \propto (1+z)^{0.54}$. The averages of $\langle 1+z \rangle$ for different components follow the analytic solution for the whole cosmic web at low redshift, but deviate from it at high redshift. For T57, the deviation is noticeable for $z \ga 1$ and becomes $\sim 35 \%$ at $z = 5$. {Note that $\langle 1+z \rangle$ is smaller (larger) if it is weighted more with the density at lower (higher) redshift along the LOS (see [\ref{eq:nz}]). In that sense, the trend of $\langle 1+z \rangle$, that is, $\langle 1+z \rangle$ for T57 and T79 smaller than that for T04 and T45, is consistent with the behavior of $f_{DM}$ for different components.}

\subsection{Rotation Measure}
\label{subsection3.2}

\begin{figure}
\begin{center}
\includegraphics[width=90mm]{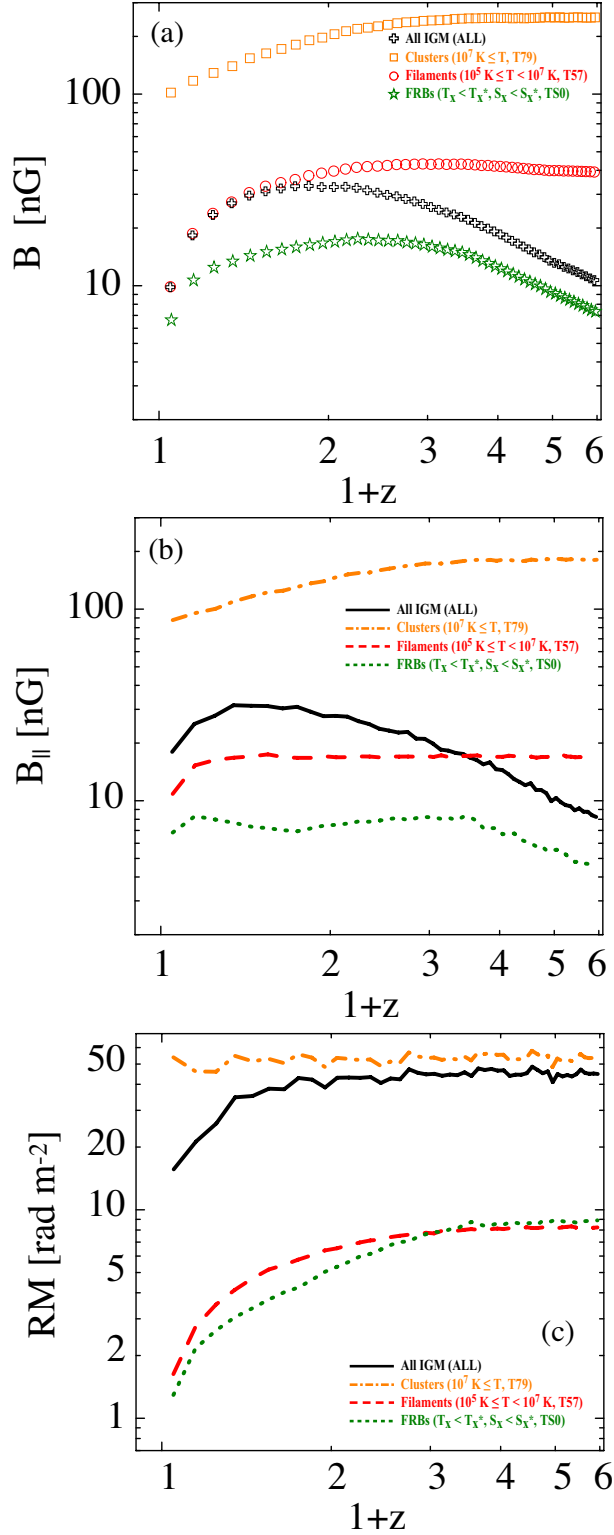}
\end{center}
\vskip -0.6cm
\caption{
(a) The density-weighted IGMF strength, $B$, (b) the density-weighted LOS
strength of the IGMF, $B_{||}$, and (c) the rotation measure, RM, each
integrated along LOSs up to a redshift $z$, for ALL (black crosses and solid
line), T79 (orange squares and dash-dotted line), T57 (red circles and
dashed line), and TS0 (green stars and dotted line). Symbols in (a) represent the average. Lines in (b) and (c) represent the rms values. $B$'s and RM for T45 and T04 are not shown (lying outside the plot range).
}\label{f3}
\end{figure}

\placefigure{f3}

We now present the rotation measure (RM) and average field strength resulting from our model IGMF. Figure \ref{f3}(a) shows the average of the density-weighted IGMF strength, $B$, integrated along LOSs for different components of the cosmic web and TS0. There are large variances in $B$ within each redshift bin (not shown for clear display), due to the highly intermittent nature of the IGMF. In our model IGMF, the average value of $B$ converges to a couple $\times~100$ nG for T79 and a few $\times ~10$ nG for T57 at large $z$ \citep[see][for further discussion of the model IGMF]{ryu08}. It is smaller for T45, a couple $\times ~0.1$ nG at large $z$. The value of $B$ should be much smaller for T04 in voids (not shown, lying outside the range of $B$ plotted).\footnote{Observational evidence suggests that the IGMF in voids has a strength $\ga 10^{-16}$ G \citep[e.g.,][]{Neronov2010, Tavecchio2010}} The average value of $B$ for T79 is larger than that for ALL, for instance, since $B$ for T79 are contributed only from the hot gas of clusters which has strongest magnetic fields in the cosmic web. The average value of $B$ for each component of the cosmic web is slightly larger at higher redshift. On the other hand, $B$ for ALL peaks at $z \simeq 0.8$ and is smaller at higher redshift, reflecting the structure formation history. The average value of $B$ for TS0, which excludes the contribution from the hot gas of clusters, is contributed mostly from the WHIM, but $B$ for TS0 is a few times smaller than that for T57 due to averaging along LOS.

Figure \ref{f3}(b) shows the rms of the density-weighted LOS strength of the IGMF, $B_{||}$, integrated along LOSs for different components of the cosmic web and for TS0. We note that the average value of $B_{||}$ is zero. The overall behavior of the rms value of $B_{||}$ is similar to that of the average value of $B$. However, for T57 the value of $B_{||,{\rm rms}}$ is close to that of $B_{\rm avg}/\sqrt{3}$, while for T79, the values of $B_{||,{\rm rms}}$ and $B_{\rm avg}$ are comparable, indicating that the model IGMF has larger variances for T79. Again, $B_{||}$ for TS0 is contributed mostly by the WHIM, but the value of $B_{||,{\rm rms}}$ for TS0 is somewhat smaller than that for T57.

Figure \ref{f3}(c) shows the rms of RM for different components of the cosmic web and for TS0. The value of ${\rm RM}_{\rm rms}$ for ALL is the same as that shown by \citet{ar11}, except that the number of LOSs used is different. With a larger gas density and stronger magnetic field, ${\rm RM}_{\rm rms}$ for T79 due to the hot gas of clusters is substantially larger than that for T57 and close to ${\rm RM}_{\rm rms}$ for ALL, as expected. Note that ${\rm RM}_{\rm rms}$ for T57 is larger than ${\rm RM}_{\rm rms}$ for ALL, because ${\rm RM}_{\rm rms}$ reflects the variance. The values of ${\rm RM}_{\rm rms}$ for T45 and T04 are much smaller (not shown, lying outside the range plotted), indicating that their contributions to the observed RM are expected to be negligible. This indicates that RM could be used to explore the magnetic field for T57, that is, in the WHIM of filaments. But for this to be feasible, the contribution due the hot gas needs to be eliminated. We suggest that this can be achieved by adopting a scheme like TS0 \citep{ar11}. Figure~\ref{f3}(c) shows that  the value of ${\rm RM}_{\rm rms}$ for TS0 is indeed close to that for T57.

\subsection{Estimation of Line-of-Sight Magnetic Field Strength}
\label{subsection3.3}

\begin{figure}
\begin{center}
\includegraphics[width=80mm]{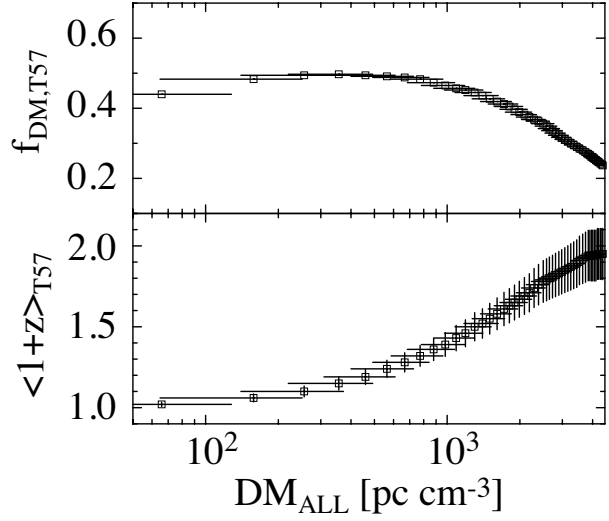}
\end{center}
\vskip -0.6cm
\caption{
The DM fraction (top) and the density-weighted redshift along sightlines (bottom) for T57, as a function of DM for all IGM (ALL). Squares and error bars mark averages and standard deviations, respectively.
}\label{f4}
\end{figure}

\placefigure{f4}

\begin{figure}
\begin{center}
\includegraphics[width=80mm]{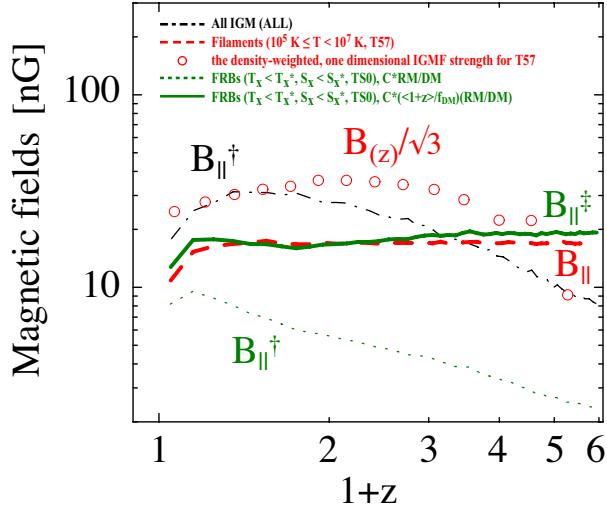}
\end{center}
\vskip -0.6cm
\caption{
The rms of the LOS IGMF estimate using values of DM and RM for TS0 ($B_{||}^\dagger$, thin green dashed line, equation [\ref{eq:RMDM}]) and the improved estimate using the DM and RM for TS0,  the average DM fraction, $f_{DM}$, and the average of the density-weighted redshift, $\langle 1+z \rangle$, for T57 ($B_{||}^\ddagger$, thick green solid line, equation [\ref{eq:Bf}]). These estimates are compared to the rms of the density-weighted LOS IGMF strength for T57 ($B_{||}$) (red dashed line, also shown in Figure \ref{f3}[b]). For reference, the rms of $B_{||}^\dagger$ using DM and RM for ALL (black dash-dotted line), and the density-weighted, one-dimensional IGMF strength at a given redshift for T57 (red circles, equation [\ref{eq:batz}]) are also shown.
}\label{f5}
\end{figure}

\placefigure{f5}

We now investigate how the values of DM and RM for the cosmic web observed towards FRBs can be used to probe the IGMF in filaments of galaxies. We point out that for $B_{||}^\dagger$ in equation~(\ref{eq:RMDM}), three limitations must be addressed. First, the component of the IGM that dominates the DM contribution changes as a function of redshift: at low redshift the main contributor is the WHIM of filaments, while at high redshift it is the gas in voids, as shown in \S\ref{subsection3.1} and Figure~\ref{f2}(a). Second, the RM is contributed mostly by the hot gas of clusters and only a fraction of it is due to the WHIM, as discussed in \S\ref{subsection3.2} and Figure~\ref{f3}. Finally, DM and RM in the cosmological context have different redshift dependences, as per equations (\ref{eq:DM}) and (\ref{eq:RM}).

These problems with equation~(\ref{eq:RMDM}) can be resolved as follows. First, instead of using the DM of the entire cosmic web to calculate the magnetic field strength, the DM only due to the WHIM should be used. {This can be achieved by replacing $DM$ with $f_{DM}DM$ in equation~(\ref{eq:RMDM}), where $f_{DM}$ is the value for T57.} Second, LOSs that avoid clusters should be chosen, via a scheme like that presented by TS0. Finally, one must include a correction for the redshift dependence, which we accomplish by substituting  {$DM/\langle 1+z \rangle$ for $DM$, where $\langle 1+z \rangle$  is the value for T57.} Based on these adjustments, we propose an improved estimate for the LOS strength of the IGMF in filaments,
\begin{equation}\label{eq:Bf}
B_{||}^\ddagger = \frac{\langle 1+z \rangle}{f_{DM}} B_{||}^\dagger
= \frac{\langle 1+z \rangle}{f_{DM}}\frac{C_{\rm D}RM}{C_{\rm R}DM}.
\end{equation}
Here, the DM and RM for TS0 are used, while for the average DM fraction, $f_{DM}$, and the average of the density-weighted redshift, $\langle 1+z \rangle$, the values for T57 gas are used (i.e., the red circle data in Fig.~\ref{f2}). 

We note that $f_{DM}$ and $\langle 1+z \rangle$ for T57 can be evaluated with relatively small errors for a given cosmology model once the DM of an FRB is known. Figure \ref{f4} shows $f_{DM}$ and $\langle 1+z \rangle$ for T57 as a function of DM for ALL in our cosmology model. This demonstrates that for the observed DMs of FRBs ($\simeq 400 - 1600$~${\rm pc~cm^{-3}}$),  errors for the evaluations of $f_{DM}$ and $\langle 1+z \rangle$ for T57 should be $\sim 10 - 20 \%$. Therefore, the redshift of an FRB does not need to be known for estimating the IGMF in filaments of galaxies using equation~(\ref{eq:Bf}), provided that any local DM and RM contributions associated with the FRB's host galaxy or immediate environment are also accounted for.

Figure \ref{f5} shows our improved estimate of the LOS strength of the IGMF in filaments, $B_{||}^\ddagger$, along with estimates without corrections, for instance, $B_{||}^\dagger$ with DM and RM for TS0 and $B_{||}^\dagger$ with DM and RM for ALL. Note that the latter is the simple estimate of magnetic field that would be derived from the observed DM and RM using equation (\ref{eq:RMDM}). These estimates derived from DM and RM are compared with the underlying density-weighted LOS strength of the IGMF for T57, $B_{||}$. The figure demonstrates that the rms of $B_{||}^\ddagger$ closely follows the rms of $B_{||}$, while the other estimates $B_{||}^\dagger$ fail to reproduce the behavior of $B_{||}$. The figure also shows $B_{(z)}/\sqrt3$ for the WHIM of T57, where
\begin{equation}\label{eq:batz}
B_{(z)} = \int n_{e,(z)} B_{(z)} dV {\bigg/} \int n_{e,(z)} dV.
\end{equation}
This demonstrates that the rms of $B_{||}^\ddagger$, which represents an integrated quantity along LOS, reproduces the density-weighted, one-dimensional IGMF strength at a given redshift within a factor of~$\sim2$.

\section{Discussion and Summary}
\label{section4}

Using the results of cosmological structure formation simulations and a model IGMF, we have calculated the dispersion measure (DM) and rotation measure (RM) induced by different components of the cosmic web, determined by integrating physical quantities along LOSs toward FRBs distributed over the redshift range $z=$ 0 -- 5. We find that the DM due to the IGM along the sightline to an FRB arises primarily in the WHIM in filaments and the gas in voids; at low redshifts, the DM due to the WHIM dominates, while at high redshifts, the DM due to the void gas is the main contributor. The DM due to the hot gas in clusters is small for most of the redshift range considered. On the other hand, with our model IGMF, RM is induced mostly by the hot gas, and the RM due to the WHIM is an order of magnitude smaller than the RM due to the hot gas.

We have then examined the proposal that the observed DMs and RMs of FRBs can be used to probe the IGMF, especially the magnetic field in galaxy filaments. Based on our results, we propose an improved estimate for the LOS strength of the IGMF in filaments, $B_{||}^\ddagger = C(\langle 1+z \rangle/f_{DM})(RM/DM)$, where $C$ is a known constant. Here, DM and RM are those observed for an FRB, provided that one only uses sightlines chosen to avoid clusters based on criteria of X-ray temperature and surface brightness, and that one excludes any contribution to DM and RM local to the FRB/its host galaxy or due to the Milky Way. $f_{DM}$ is the fraction of intergalactic DM due the WHIM, and $\langle 1+z \rangle$ is the redshift weighted by the WHIM gas density (equation [\ref{eq:nz}]). {The values of $f_{DM}$ and $\langle 1+z \rangle$ can be evaluated for a given cosmology model if the value of DM is known; hence we do not need to know the redshift of the FRB.} We have shown that with our model cosmology and IGMF, the rms of $B_{||}^\ddagger$ is almost identical to the rms of the density-weighted LOS strength of the IGMF in filaments. Our work suggests that if enough DMs and RMs of FRBs can be determined, the strength of the IGMF in galaxy filaments could then be estimated.

There are uncertainties in the estimation. First, we have used DM as an indicator of the redshift of an FRB. Although DM is a tight function of redshift, there is a variance. The variance introduces an uncertainty in deriving the redshift from the observed DM, at an estimated level $\delta z \sim 0.2$ at a 1$\sigma$ level. The uncertainty in redshift propagates into the evaluations of $f_{DM}$ and $\langle 1+z \rangle$ for the WHIM. The uncertainties in the evaluations of $f_{DM}$ and $\langle 1+z \rangle$ are estimated to be $\sim 10 - 20 \%$. Overall, the uncertainty in $B_{||}^\ddagger$ should be at most a few to several $\times ~10 \%$. In future, if host galaxies of FRBs are identified and their redshifts are determined by follow-up observations such as line measurements, $f_{DM}$ and $\langle 1+z \rangle$ for the WHIM can be directly evaluated from the redshift instead of from the DM, reducing the uncertainties in the field strength estimate.

There of course are other, possibly larger, uncertainties that we have not accounted for here. Recent studies have noted that the DM contributions of FRB host galaxies \citep[e.g.,][]{con15,ma15b,kon15} and of the foreground Milky Way and Local Supercluster \citep[see][]{dolag15} could all be significant. Likewise, the RM contributed by the FRB's immediate environment, host galaxies, intervening galaxies along the LOS and the Milky Way could all be larger than the RM due to the WHIM \citep[see, e.g.,][]{bdlk13,agr14}. The above additional contributions could be partly incorporated through further modeling, statistical approaches and Faraday synthesis \citep[e.g.,][]{agr14,aktr14}. However, this will inevitably introduce additional uncertainties in $B_{||}^\ddagger$. 

The number of FRBs needed to reliably estimate the IGMF in galaxy filaments would depend on such unknown foreground and host contributions as well as the cosmic variance. The estimation of the number is thus beyond the scope of this paper. Further work using numerical simulations is needed to establish how many FRB detections and how wide a survey area are needed to overcome involved uncertainties.

Finally, it would be interesting to apply our results to observed data. As
mentioned in \S\ref{section1}, the linearly polarized FRB~110523 has DM $=
623$ ${\rm pc~cm^{-3}}$ and RM $= -186$ ${\rm rad~m^{-2}}$ \citep{ma15b}.
The observed amplitude of RM is too large to ascribe to magnetic fields in
filaments. The authors suggest that the RM of the FRB could
be due to magnetic fields in the vicinity of the source itself or within the
host galaxy \citep[see also Fig.~14 of][]{bdlk13}. However, the RM could be also
due to magnetic fields in intervening and/or host galaxy clusters.
Although our equation~(\ref{eq:Bf}) was presented for the WHIM (T57), it can
be also applied to the hot gas in clusters (T79). If we apply
equation~(\ref{eq:Bf}) to the observed values of RM and DM for FRB~110523,
along with $f_{\rm DM}$ and $\langle 1+z \rangle$ for T79, we obtain
$B_{||}^\ddagger \sim 7$~$\mu$G, which is somewhat strong, but possible for
magnetic fields in clusters \citep[e.g.,][]{tfa02}.
{A similar procedure could be applied to FRB 150418 with DM
$= 776.2 \pm 0.5$ ${\rm pc~cm^{-3}}$ and RM $= +36 \pm 52$ ${\rm rad~m^{-2}}$
\citep{kea16}.
The quoted RM could be consistent with that due to galaxy filaments or
clusters (see Fig.~3[c]), but the uncertainty is too large to make any
conclusive statement.
This emphasizes the need for accurate measurements of RMs for the estimation of the IGMF
with FRBs.}

\acknowledgements

T.~A. was supported by JSPS KAKENHI Grant Numbers 15K17614 and 15H03639. D.~R. was supported by the National Research Foundation of Korea through NRF-2014M1A7A1A03029872. B.~M.~G. acknowledges the support of the Australian Research Council through grant FL100100114. The Dunlap Institute is funded through an endowment established by the David Dunlap family and the University of Toronto.

\end{document}